\documentclass[%
 preprint,
 amsmath,amssymb,
 aps, physrev,
]{revtex4-2}

\usepackage{graphicx}
\usepackage{dcolumn}
\usepackage{bm}

\begin{document}

\title{\textbf{The Complete Set of Solutions for the Equations of Temporal Coupled Mode Theory} 
}%

\author{Matthew Beutel}
 \affiliation{Department of Applied Physics, Stanford University, Stanford, California 94305, USA.}
\author{Ming Zhou}
\author{Shanhui Fan}
 \email{Contact author: shanhui@stanford.edu}
\affiliation{%
 Ginzton Laboratory and Department of Electrical Engineering, Stanford University, Stanford, California 94305, USA.
}%
\date{\today}

\begin{abstract}
Temporal coupled mode theory (TCMT) provides a compact description of resonant transport in optical systems in terms of a small set of physically transparent parameters, including the resonant coupling coefficients and direct background scattering matrix. In the formalism of TCMT, the background scattering matrix and the coupling vectors are constrained by time reversal symmetry, reciprocity, and energy conservation. While these constraints are well established, the general solution under these is only known for a few cases, in which spatial symmetry simplifies the problem. In this paper, we derive the complete set of solutions for the equations of temporal coupled mode theory that describes lossless single-mode or multi-mode resonators coupled to an arbitrary number of ports. Our derivation utilizes an explicit Takagi decomposition of the symmetric unitary background scattering matrix. We provide a comparison with numerical simulations for both the single and multimode cases.
\end{abstract}

\maketitle


\section{\label{sec:level1}Introduction}

A resonant optical device consists of a resonator supporting one or multiple resonances, coupling with multiple ports for input and output. In integrated photonics, the ports consist of on-chip waveguides, and the resonators can be micro-ring resonators \cite{xu2005micrometre} or photonic crystal resonators \cite{fan1998channel}. For metasurfaces, the ports are plane waves from various directions, and the resonators can be formed with the use of either Mie resonances in individual dielectric or metallic resonators \cite{kuznetsov2016optically}, or through the creation of guided resonances in photonic crystal slabs \cite{fan2002analysis}. In all these systems, the presence of the resonator significantly affects the transport properties of light between the ports, leading to applications such as optical filters, sensors and switches.  

The temporal coupled mode theory formalism provides a general framework in the study of transport processes in resonant optical devices \cite{haus1984waves,fan2003temporal,suh2004temporal,PhysRevA.99.033839}. The formalism describes the transport process in terms of the interference of two pathways \cite{fan2003temporal}. The first pathway is a resonant process in which input waves from the ports excite the resonance, the resonances then decay back to the ports forming the output waves. This pathway is characterized by a set of complex amplitude decay rates of the resonances to the various ports. The second pathway is a direct process where the input waves are directly scattered into the output waves without exciting the resonance. This pathway is described in terms of a background scattering matrix. The parameters that describe these two pathways are related by general arguments such as energy conservations, time-reversal symmetry, and reciprocity.

The temporal coupled mode theory formalism thus builds in the general physical constraints and describes the resonant systems in terms of a few parameters with well-defined physical interpretations. This formalism is particularly well-suited to generate physical insights and to develop design criteria for applications. It has been applied to describe resonant filters \cite{manolatou2002coupling,Bykov:15}, perfect absorbers \cite{piper2014total, piper_total_2014,Ming:17,tran_graphene-assisted_2019}, solar absorbers \cite{yu2010fundamental,yu2012thermodynamic}, anti-reflectors \cite{wang2014condition}, thermal emitters \cite{zhu2013temporal,PhysRevApplied.19.034037,PhysRevX.11.021050}, optical antennas \cite{ruan2010temporal}, optical sensing \cite{Chen:20,park_critically_2021, ren_wavelength-multiplexed_2022, Wang:19}, inverse design \cite{10.1021/acsphotonics.1c00100}, as well as active and nonlinear devices \cite{yanik2003high,PhysRevA.89.053839, PhysRevA.83.033834, PhysRevA.89.053839}. Moreover, this formalism is connected to the theory of quasi-normal mode \cite{wu2024designing, PhysRevX.7.021035,PhysRevResearch.3.033228,bulgakov_nonlinear_2019}, which allows many of the parameters in temporal coupled mode theory to be computed from first principles efficiently. The formalism of temporal coupled mode theory has also been generalized to spatial and spatial-temporal domains \cite{overvig_spatio-temporal_2024,Ruan:15,Lou:16}.

In the temporal coupled mode theory formalism, the decay rates of the resonators into various ports are constrained by the background scattering matrix in an equation that was first derived in Ref. \cite{fan2003temporal}.  The general solution of this equation however is not known except for a few specific cases, where the structure possesses spatial symmetries. For the general case where the structure may not possess any symmetry, Ref. \cite{wang2013fundamental} derived a bound regarding the ratio of the decay rates into various ports as constrained by the background scattering matrix. This bound is subsequently shown to be related to a time-reversal symmetry bound of the Maxwell’s equations \cite{sounas2017time}. More recently, for arbitrary numbers of modes and ports, constraints on both the magnitudes and phases of the resonant coupling coefficients have been generalized \cite{wang2024timereversalsymmetryboundstemporal}. 

As a step further, in this paper we provide a general solution of the decay rates in the temporal coupled mode theory as constraint by the background scattering matrix. The solution is based on an explicit construction of a Takagi decomposition of the background scattering matrix. In the single-mode case,  we show that the decay rates to the eigen-channels of the background scattering matrix can be arbitrarily chosen. These decay rates to the eigen-channels then completely parameterize all possible solutions. For the multimode cases, these decay rates from different modes are in addition subject to orthogonality constraints. As an illustration, we consider the two-port cases with one or two modes in detail. We also provide numerical validations of the theory.

\section{\label{sec:level1}Theory }
\subsection{Single Mode Case}
\begin{figure}
    \centering
    \includegraphics[width=1\linewidth]{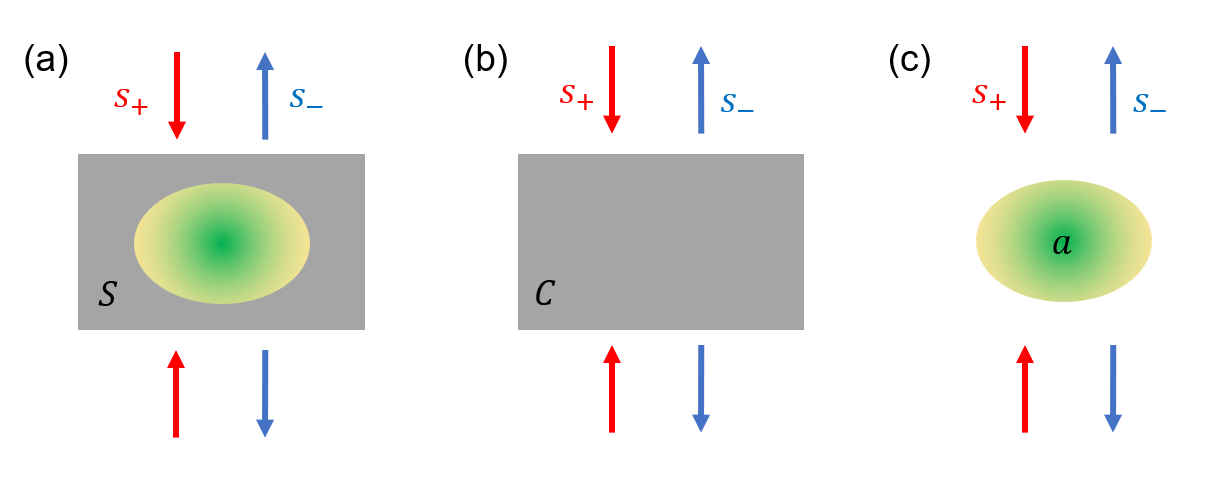}
    \caption{Illustration of the temporal coupled mode theory formalism. The aim of the theory (a) is to compute the scattering matrix (S) of a system supporting resonances. The system can be decomposed into (b) a background scattering pathway and (c) a resonant pathway. }
    \label{fig:2port}
\end{figure}

We start with a brief review of the temporal coupled mode theory for a single-mode resonator as developed in Ref. \cite{fan2003temporal}.  With the assumption of energy conservation and reciprocity, for a lossless single-mode resonator coupling to \emph{n} input/output ports, its property can be described as:

\begin{equation}
\frac{da}{dt} = (i\omega_0 - \gamma)\,a + \kappa^{T} s_{+}
\label{eq:da/dt}
\end{equation}

\begin{equation}
s_{-} = C\,s_{+} + d\,a .
\label{eq:sminus}
\end{equation}
Here, $a$ is the amplitude of the resonance with $\lvert a\rvert^{2}$ corresponding to the energy of the resonance.
$\omega_{0}$ and $\gamma$ are the resonant frequency and the total decay rate, respectively.
$s_{+} = (s_{1+}, \ldots, s_{n+})^{T}$ and $s_{-} = (s_{1-}, \ldots, s_{n-})^{T}$, with $s_{i+}$ and $s_{i-}$ being the input (output) wave amplitudes at each of the $i$th port.
$\lvert s_{i+}\rvert^{2}$ ($\lvert s_{i-}\rvert^{2}$) is the input (output) power at the $i$th port (Fig. \ref{fig:2port}).
$\kappa^{T} = (\kappa_{1}, \ldots, \kappa_{n})$, where $\kappa_{i}$ is the input coupling constant from the $i$th port to the resonator.
$d^{T} = (d_{1}, \ldots, d_{n})$, where $d_{i}$ is the output coupling constant from the resonator to the $i$th port.
$C$ is the background scattering matrix and describes the contribution of non-resonant processes through the resonator.
For our system which is reciprocal and lossless, we assume that the background scattering matrix is unitary and symmetric.

For the temporal coupled mode theory formalism of Eqs. \eqref{eq:da/dt} and \eqref{eq:sminus}, using
energy conservation and reciprocity, one can show that \cite{fan2003temporal}
\begin{equation}
d^{\dagger} d = 2\gamma
\label{eq:d_dagger_d}
\end{equation}
\begin{equation}
\kappa = d
\label{eq:kappa_equals_d}
\end{equation}
\begin{equation}
C\, d^{*} = -d
\label{eq:C_dstar_equals_minus_d}
\end{equation}
Equation~\eqref{eq:d_dagger_d} relates the total decay rate of the resonator
to the output coupling constants and is a direct consequence of energy
conservation \cite{haus1984waves}. Equation~\eqref{eq:kappa_equals_d} equates the input and output
coupling constants and is a result of reciprocity \cite{PhysRevA.99.033839}. Equation~\eqref{eq:C_dstar_equals_minus_d}
shows that there are non-trivial constraints on the output coupling constants
as set by the background scattering matrix \emph{C}.

Below we construct all possible solutions to Eq.~\eqref{eq:C_dstar_equals_minus_d}
with Takagi decomposition \cite{chebotarev2014singular}. Takagi decomposition, when restricted
to the matrix $C$, which is both unitary and complex symmetric, states that
there exists a unitary matrix $V$ such that
\begin{equation}
C = VV^{T}.
\label{eq:takagi_C}
\end{equation}
Combining Eqs.~\eqref{eq:C_dstar_equals_minus_d} and~\eqref{eq:takagi_C}, we have
\begin{equation}
(V^{\dagger} d)^{*} = -(V^{\dagger} d).
\label{eq:Vdaggerd_constraint}
\end{equation}
Therefore, we can set $V^{\dagger} d = i d_{0}$, where $d_{0}$ is a
$real$ vector satisfying
\begin{equation}
d_{0}^{\dagger} d_{0} = 2\gamma.
\label{eq:d0_norm}
\end{equation}
The solution of $d$ is then
\begin{equation}
d = i V d_{0}.
\label{eq:d_solution_basic}
\end{equation}
The matrix $V$ can be explicitly constructed as follows: Since $C$ is a unitary
matrix, it is unitarily diagonalizable \cite{zhang2011matrix}, i.e.
\begin{equation}
C = W e^{i\Theta} W^{\dagger},
\label{eq:C_spectral}
\end{equation}
where
\begin{equation}
\Theta = \mathrm{diag}\{\theta_{1},\ldots,\theta_{n}\},
\label{eq:Theta_def}
\end{equation}
with all the $\theta_i$'s being real. Moreover, since $C$ is a symmetric
matrix, $W$ can be chosen to be real, i.e.\ $W = W^{*}$. Therefore, we can choose
\begin{equation}
V = W e^{i\Theta/2}
\label{eq:V_choice}
\end{equation}
in the Takagi decomposition of $C$. From Eq.~\eqref{eq:d_solution_basic}, we now have
\begin{equation}
d = i W e^{i\Theta/2} d_{0}.
\label{eq:d_final}
\end{equation}
Eq.~\eqref{eq:d_final} is the main result of this section. It parameterizes all
possible solutions of Eq.~\eqref{eq:C_dstar_equals_minus_d}.

We elucidate the physical meaning of $d_{0}$. First of all, from
Eq.~\eqref{eq:C_spectral}, we have
\begin{equation}
C\, w_{j} = e^{i\theta_{j}} w_{j},
\label{eq:C_eig}
\end{equation}
where $w_{j}$ is the $j^{\mathrm{th}}$ column of the matrix $W$. Thus, $w_{j}$
is an eigenchannel of the background scattering matrix. For an incident wave
having the form of $w_{j}$, the output wave that results from the scattering process, as defined by the background scattering matrix $C$, has the same form,
acquiring only an overall phase shift $e^{i\theta_{j}}$. From
Eq.~\eqref{eq:d_final}, we then have
\begin{equation}
\left|d_{0,j}\right| = \left|w_{j}^{\dagger} d\right|.
\label{eq:d0j_mag}
\end{equation}
Thus, $|d_{0,j}|$ is the magnitude of the output coupling strength from the
resonator to the $j^{\mathrm{th}}$ eigenchannel. Eqs.~\eqref{eq:d_final}
and~\eqref{eq:d0j_mag} indicate that to obtain the solution of
Eq.~\eqref{eq:C_dstar_equals_minus_d}, the coupling constant to each
eigenchannel can be set to be an arbitrary real number, subject only to the
total decay rate constraint of Eq.~\eqref{eq:d0_norm}. The decay rate to each
output channel is then parameterized by those to the eigenchannels, in a form
as expressed in Eq.~\eqref{eq:d_final}.

We now solve for the transport properties of the system as described by the
temporal coupled mode theory of Eqs. \eqref{eq:da/dt}-\eqref{eq:C_dstar_equals_minus_d}. Following Ref. \cite{fan2003temporal}, for
light incident at a frequency $\omega$, the system's scattering matrix, which
relates the output wave amplitudes to the input wave amplitudes, is given by:
\begin{equation}
S = C + \frac{d d^{T}}{i(\omega-\omega_{0})+\gamma}.
\label{eq:S_general}
\end{equation}
Using Eq.~\eqref{eq:d_final}, we then have
\begin{equation}
S = W e^{i\Theta/2}
\left[I - \frac{d_{0} d_{0}^{T}}{i(\omega-\omega_{0})+\gamma}\right]
e^{i\Theta/2} W^{T}.
\label{eq:S_W_form}
\end{equation}
Eq.~\eqref{eq:S_W_form}, with $d_0$ being real and subject to the constraint of $d_0^T d_0 = 2 \gamma$,  provides $all$ possible solutions of the
scattering matrix of the single-resonance system for a given background
scattering matrix $C$.

We illustrate the discussion above by considering the two-port case. In the
two-port case, there exists a reference plane choice such that the background
scattering matrix is of the form:
\begin{equation}
C =
\begin{pmatrix}
r & i t\\
i t & r
\end{pmatrix}
=
\begin{pmatrix}
\cos\theta & i\sin\theta\\
i\sin\theta & \cos\theta
\end{pmatrix},
\label{eq:C_twoport}
\end{equation}
where, without loss of generality, we assume $r>0$. $t=\sqrt{1-r^{2}}$. We
perform a Takagi decomposition $C=V V^{T}$ following the procedure above where
we diagonalize the matrix $C$ first, to obtain
\begin{equation}
V=\frac{1}{\sqrt{2}}
\begin{pmatrix}
e^{i\theta/2} & e^{-i\theta/2}\\
e^{i\theta/2} & -e^{-i\theta/2}
\end{pmatrix}.
\label{eq:V_twoport}
\end{equation}
Let
\begin{equation}
d_{0}=\sqrt{2\gamma}
\begin{pmatrix}
\cos\alpha\\
\sin\alpha
\end{pmatrix}.
\label{eq:d0_twoport}
\end{equation}
Then
\begin{equation}
d=i V d_{0}
= i\sqrt{\gamma}
\begin{pmatrix}
e^{i\theta/2}\cos\alpha + e^{-i\theta/2}\sin\alpha\\
e^{i\theta/2}\cos\alpha - e^{-i\theta/2}\sin\alpha
\end{pmatrix}.
\label{eq:d_twoport}
\end{equation}
Using Eq.~\eqref{eq:S_general}, the scattering matrix for such two-port system is
\begin{equation}
S=
\begin{pmatrix}
r & i t\\
i t & r
\end{pmatrix}
+\frac{\gamma}{i(\omega-\omega_{0})+\gamma}
\begin{pmatrix}
-r-\sin 2\alpha - i t\cos 2\alpha & -r\cos 2\alpha - i t\\
-r\cos 2\alpha - i t & -r+\sin 2\alpha - i t\cos 2\alpha
\end{pmatrix},
\label{eq:S_twoport}
\end{equation}
where $\alpha$ can be arbitrarily chosen.
From Eq.~\eqref{eq:S_twoport}, we can also show that
\begin{equation}
\left|d_{1,2}\right|^{2} = (1 \pm r \sin 2\alpha)\gamma
\label{eq:d12_mag_sq}
\end{equation}
Thus, in case where $r>0$,
\begin{equation}
\frac{1-r}{1+r} \le \frac{|d_{1}|^{2}}{|d_{2}|^{2}} \le \frac{1+r}{1-r}.
\label{eq:d1d2_ratio_bounds}
\end{equation}
Eq. \eqref{eq:d1d2_ratio_bounds} was first obtained in Ref. \cite{wang2013fundamental} and shown to be connected to some
of the general properties of electromagnetic Green's function in
Ref. \cite{sounas2017time}. Our formalism is consistent with Ref. \cite{wang2013fundamental}. The form
of scattering matrix as shown in Eq.~\eqref{eq:S_twoport} has not been
previously reported.

\begin{figure}
    \centering
    \includegraphics[width=0.65\linewidth]{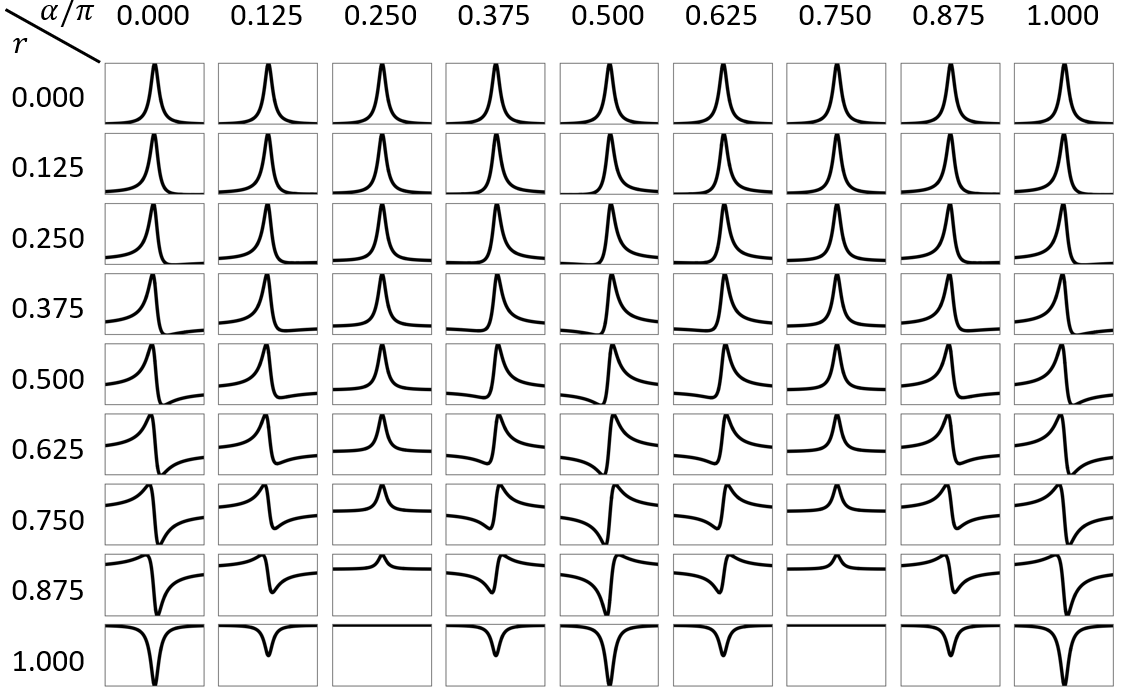}
    \caption{Intensity reflection spectra in a two-port system as one varies the background amplitude reflection coefficient $r$, and the factor $\alpha$ that controls the ratio of the decay rates into the two ports. In each panel, the vertical axis is the intensity reflection coefficient with values ranging from 0 to 1. The horizontal axis is the normalized frequency detuning of $(\omega - \omega_0)/\gamma$ with values ranging from -10 to 10. 
}
    \label{fig:1modespectra}
\end{figure}

Eq.~\eqref{eq:d_twoport} exhibits a rich set of transport behaviors, as
illustrated in Figure~2, where we plot the reflection spectra of the system.
In general, there are two types of line shapes: a symmetric type, or a Fano
line shape that is asymmetric. Below, we consider a few special cases. The
symmetric line shapes occur at: \\
\noindent\textbf{Case 1:} $r=0$. In this case, the reflection reaches unity on
resonance. We have $|d_{1}|=|d_{2}|$ independent of the choice of $\alpha$. The
reflection line shape is also independent of $\alpha$. This case can be
realized with a waveguide side-coupled to a standing-wave cavity, as can be
realized in photonic crystal systems \cite{fan2002sharp}. \\
\noindent\textbf{Case 2:} $r=1$. This case corresponds to a high-$Q$
Fabry--Perot cavity for which the background reflection can be approximated as
unity. When $\alpha = N\pi/2$, with $N\in\mathbb{Z}$, we have
$|d_{1}|=|d_{2}|$. The Fabry--Perot cavity is symmetric. The reflection
vanishes at resonance. At other values of $\alpha$, the Fabry--Perot cavity is
not symmetric. The reflection spectrum remains symmetric, but the reflection no longer reaches zero on resonance. The example spectra of both cases are shown in Fig. \ref{fig:1modespectra}. \\
\noindent\textbf{Case 3:} $\alpha=\pi/4$ or $\alpha=3\pi/4$. The reflection
line-shapes consist of a \emph{symmetric} reflection peak, superimposed on a
background. In this case, the ratio $|d_{1}|^{2}/|d_{2}|^{2}$ reaches the
bound described in Eq. \eqref{eq:d1d2_ratio_bounds}. \\
\indent For all other cases of $r$ (i.e., $r\neq 0$ and
$r\neq 1$), the reflection spectrum is asymmetric. For all these cases, the
reflections maximize at unity. The reflection, aside from the cases where $\alpha=N\pi/2$ where $N$ is an integer, on the other hand, is not zero. Thus, for most values of $\alpha$, the system supports perfect reflection but not perfect transmission. Also, for a given background reflection coefficient $r$, we note that the reflection spectra at
$\alpha=\pi/4\pm\delta$ (or $\alpha=3\pi/4\pm\delta$) are mirror images of
each other with respect to the frequency $\omega = \omega_0$.

\subsection{Multimode Case}
The single-resonance construction from Eqs. \eqref{eq:da/dt} and  \eqref{eq:sminus} naturally extends to the case of multiple resonant modes. For a multimode system consisting of $m$ resonances coupled to $n$ external ports, its dynamics is described by \cite{suh2004temporal} 


\begin{equation}
\frac{d{a}}{dt}=\left(i{\Omega}-{\Gamma}\right){a}+D^{T}{s}_{+}
\label{eq:multimode_a}
\end{equation}

\begin{equation}
{s}_{-}=C{s}_{+}+D{a}.
\label{eq:multimode_s}
\end{equation}
Here $a \in C^m$ with the $i$-th component of $a$ corresponding to the modal amplitude in the $i$-th mode. $\Omega$ and $\Gamma$ are both $m \times m$ Hermitian matrices. $\Omega$ describes the resonant frequencies and the Hermitian part of the coupling among the $m$ modes. $\Gamma$ describes the radiative decay due to the leakage into the ports. Both $\Omega$ and $\Gamma$ are real and symmetric. $\Gamma$ is in addition positive semi-definite. $s_-$ and $s_+$ are both $\in C^n$, with the $i$-th components of each vector corresponding to the output and input amplitudes in the $i$-th port, respectively. The matrix $D \in C^{(n \times m)}$ describes the process of leakage of the modes into the ports, with $|d_{ij}|^2$ corresponding to the energy leakage rate of the $j$-th mode into the $i$-th port. $C$ is an $n \times n$ symmetric unitary matrix that describes the background scattering process. Similar to the one-mode case, it was shown in Ref. \cite{suh2004temporal}  that due to the considerations of energy conservation and reciprocity, the following constraints are satisfied:
\begin{equation}
D^\dagger D = 2 \Gamma
\label{eq:gammad}
\end{equation}
\begin{equation}
    C D^* = - D.
    \label{eq:CDmatrix}
\end{equation}
The resulting scattering matrix takes the form:

\begin{equation}
S = C + D[i(\omega I - \Omega) + \Gamma]^{-1}D^T.
\label{eq:S_2mode}
\end{equation}

Similar to the one-mode case, here we aim to find all possible solutions of $D$ that satisfy Eqs. \eqref{eq:gammad} and \eqref{eq:CDmatrix}. We start with Eq. \eqref{eq:CDmatrix}. Again, performing the Takagi decomposition of $C = VV^T$, where $V$ is a unitary matrix, and substituting it into Eq. \eqref{eq:CDmatrix}, we get
\begin{equation}
    (V^\dagger D)^* = -V^\dagger D.
\end{equation}
Thus we can choose,
\begin{equation}
    D = iVD_0, \ D_0 \in \mathbb{R}^{n \times m},
    \label{eq:DVmatrix}
\end{equation}
where $D_0$ is real. Inserting Eq. \eqref{eq:DVmatrix} into Eq. \eqref{eq:gammad}, we have
\begin{equation}
    2\Gamma = D_0^TD_0.
    \label{eq:gammad0}
\end{equation}
Eq. \eqref{eq:gammad0} is consistent with the constraint stated above that $\Gamma$ is real symmetric and positive semidefinite. Moreover,
\begin{equation}
  \text{ rank}(\Gamma) = \text{rank}(D_0) \leq \text{min}(n,m)
\end{equation}
Taking $r = \text{rank}(\Gamma)$, we can perform a singular value decomposition (SVD) on $D_0$
\begin{equation}
    D_0 = P_r\Sigma_r Q_r^T.
    \label{eq:D0SVD}
\end{equation}
 Since $D_0$ is real, the SVD can be performed with $P_r, Q_r$ being real orthonormal matrices. This factorization has a simple physical explanation. Here, the columns of $P_r$ correspond to a set of orthogonal radiation patterns in the ports. Each of these radiation patterns is generated from a specific modal pattern as described by a column of $Q_r$. And these modal patterns are also orthogonal to each other.  The strength of coupling between a modal pattern and its corresponding radiation pattern in the port is described by a diagonal element of $\Sigma_r$. To satisfy the orthogonality requirements for the columns of matrices $P_r$ and $Q_r$,  we take 
\begin{equation}
    P_r \in \text{St}(n,r), \ Q_r \in \text{St}(m,r).
\end{equation}
Here $\text{St}(k,r)=\{X\in\mathbb R^{k\times r}:X^TX=I_r\}$ denotes the Stiefel manifold of ordered orthonormal $r$-frames in $\mathbb R^k$ \cite{James_1977,doi:10.1137/S0895479895290954}. Substituting the result of Eq. \eqref{eq:D0SVD} into Eq. \eqref{eq:gammad0} we get
\begin{equation}
    2\Gamma = Q_r\Sigma_r^2Q_r^T.
    \label{eq:gammaSVD}
\end{equation}
Therefore, the columns of  $Q_r$ are also the eigenvectors of $\Gamma$ that correspond to non-zero eigenvalues, which are the diagonal elements of $\Sigma^2/2$. \\
\indent We can therefore construct all solutions of the $D$ matrix in the following procedure. We start by diagonalizing the $\Gamma$ matrix (Eq. \eqref{eq:gammaSVD}) to obtain its non-zero eigenvalues and the corresponding eigenvectors to form $\Sigma_r$ and $Q_r$. Then, one can choose any $P_r \in St(n,r)$ to construct $D_0$ using Eq. \eqref{eq:D0SVD} and then $D$ using Eq. \eqref{eq:DVmatrix}. The resulting $D$ matrix thus simultaneously satisfies Eqs. \eqref{eq:gammad} and \eqref{eq:CDmatrix}. 
This procedure allows one to construct all possible solutions of Eqs. \eqref{eq:gammad} and \eqref{eq:CDmatrix}. Following this construction, we can then compute the scattering matrix elements using $C, D, \Omega, \Gamma$ in Eq. \eqref{eq:S_2mode} allowing us to compute $S$ for the multi-mode, multi-port case:
\begin{equation}
S(\omega)
=
V\left[
I_n
-
P_r\Sigma_r Q_r^{T}
\left(
i(\omega I_m-\Omega)+\Gamma
\right)^{-1}
Q_r\Sigma_r P_r^{T}
\right]V^{T}.
\label{eq:stiefelS}
\end{equation}
\indent In the case of $n < m$, $\text{rank}(\Gamma) = r \leq n < m$. In such a case, the selected $r$ orthonormal modal combinations of $Q_r$ no longer span $\mathbb{R}^m$. Because of this, there must then be $m-r$ mutually orthonormal "dark" modal combinations that do not directly contribute to any amplitude in the port space, and are not represented in $Q_r$. These modal combinations, however, can contribute to the transport process due to the coupling matrix $\Omega$.  

Here we briefly connect and contrast the single-mode and multi-mode cases. The multimode construction is a direct extension of the single-mode result: each column of $(D_0)$ is the real coupling vector of one resonant mode expressed in the background-eigenchannel coordinate system, just as $(d_0)$ is in the single-mode case. In the multimode case, however different columns of $D_0$ cannot generally be chosen independently, because energy conservation imposes the collective constraint $(D_0^T D_0=2\Gamma)$: their norms determine the individual radiative decay rates, while their mutual inner products determine the off-diagonal decay couplings between modes.

\indent To illustrate the theory above, we consider the two-mode, two-port case. In this case, we can express $\Omega$ as:
\begin{equation}
\Omega =
\begin{pmatrix}
\omega_0 - \dfrac{\Delta\omega}{2} & \kappa \\
\kappa & \omega_0 + \dfrac{\Delta\omega}{2}
\end{pmatrix},
\label{eq:2portomega}
\end{equation}
 Here, the coupling constant $\kappa$ is real since $\Omega$ is Hermitian and we assume a reciprocal system. $\Delta \omega$ is the frequency splitting between the two modes in the absence of coupling. 
We use the same background scattering matrix $C$ and its Takagi factorization in Eqs. \eqref{eq:C_twoport},\eqref{eq:V_twoport}.
Let us take  
\begin{equation}
    \Sigma_2^2=
2\begin{pmatrix}
\gamma_1 & 0 \\
0 & \gamma_2
\end{pmatrix},
\end{equation}
with $\gamma_1, \gamma_2 \geq 0$. Then, following Eq. \eqref{eq:gammaSVD}, we can construct a generic non-diagonal $\Gamma$ as
\begin{equation}
    2\Gamma = Q_2(\zeta)\Sigma_2^2Q^T_2(\zeta)
\end{equation}
where $Q_2(\zeta)$ is defined as:
\begin{equation}
Q_2(\zeta) =
    \begin{pmatrix}
\cos\zeta& \mp\sin\zeta \\
\sin\zeta& \pm\cos\zeta
\end{pmatrix} \in \text{St}(2,2).
\label{eq:2mQ}
\end{equation}
It follows then that, using Eq. \eqref{eq:D0SVD} that by constructing a $P_2(\beta)$
\begin{equation}
P_2(\beta) =
    \begin{pmatrix}
\cos\beta & \mp\sin\beta \\
\sin\beta & \pm\cos\beta
\end{pmatrix} \in \text{St}(2,2),
\label{eq:p_r_2mode}
\end{equation}
$D_0$ can then be defined as:
\begin{equation}
    D_0 = \sqrt{2}P_2(\beta)\begin{pmatrix}
        \sqrt{\gamma_1} & 0 \\
        0 &
        \sqrt{\gamma_2}
    \end{pmatrix}Q^T_2(\zeta).
    \label{eq:p2full}
\end{equation}
Here, $\beta,\zeta$, together with a sign choice, parametrizes the elements in St(2,2). Geometrically, $\beta,\zeta$ are rotational angles of the 2-frame in a two-dimensional space. Under this construction, the scattering matrix from Eq. \eqref{eq:stiefelS} can now be expressed as:
\begin{equation}
S(\omega)
=
V\left[
I_2
-
P_2(\beta)\Sigma_2 Q_2(\zeta)^{T}
\left(
i(\omega I_2-\Omega)+\Gamma
\right)^{-1}
Q_2(\zeta)\Sigma_2 P_2^{T}(\beta)
\right]V^{T}.
\end{equation}

For our numerical examples we will assume a diagonal $\Gamma$. Since $\Gamma$ is Hermitian, one can always select a modal basis that diagonalizes it. Under this assumption, we restrict the solution space of $Q_2$ in order to recover a diagonal $\Gamma$, with the desired elements. Without loss of generality, we choose $Q_2$ to be the identity matrix
\begin{equation}
    Q_2 =
     \begin{pmatrix}
1 & 0 \\
0 & 1
\end{pmatrix}. 
\end{equation}
With such a choice,
we can then construct $D_0$ using Eq. \eqref{eq:p2full}.
\begin{equation}
    D_0 = \sqrt{2}P_2(\beta)\begin{pmatrix}
        \sqrt{\gamma_1} & 0 \\
        0 &
        \sqrt{\gamma_2}
    \end{pmatrix}
\end{equation}
Further, without loss of generality, the sign in Eq. \eqref{eq:p_r_2mode} can be fixed, with this degree of freedom being absorbed by the flexibility of the sign of $\kappa$ in $\Omega$, giving:
\begin{equation}
    P_2(\beta) =
    \begin{pmatrix}
\cos\beta & -\sin\beta \\
\sin\beta & +\cos\beta
\end{pmatrix}.
\label{eq:p_2_final}
\end{equation}
Therefore, using Eq. \eqref{eq:stiefelS}, we obtain the scattering matrix for the two-port, two-mode case with diagonal $\Gamma$:
\begin{equation}
    S(\omega)
=
V\left[
I
-2P_2(\beta)\Gamma^{1/2}
\left(
i(\omega I-\Omega)+\Gamma
\right)^{-1}
\Gamma^{1/2}P_2^{T}(\beta)
\right]V^{T}.
\label{eq:2modeS}
\end{equation}

\begin{figure}
    \centering
    \includegraphics[width=0.75\linewidth]{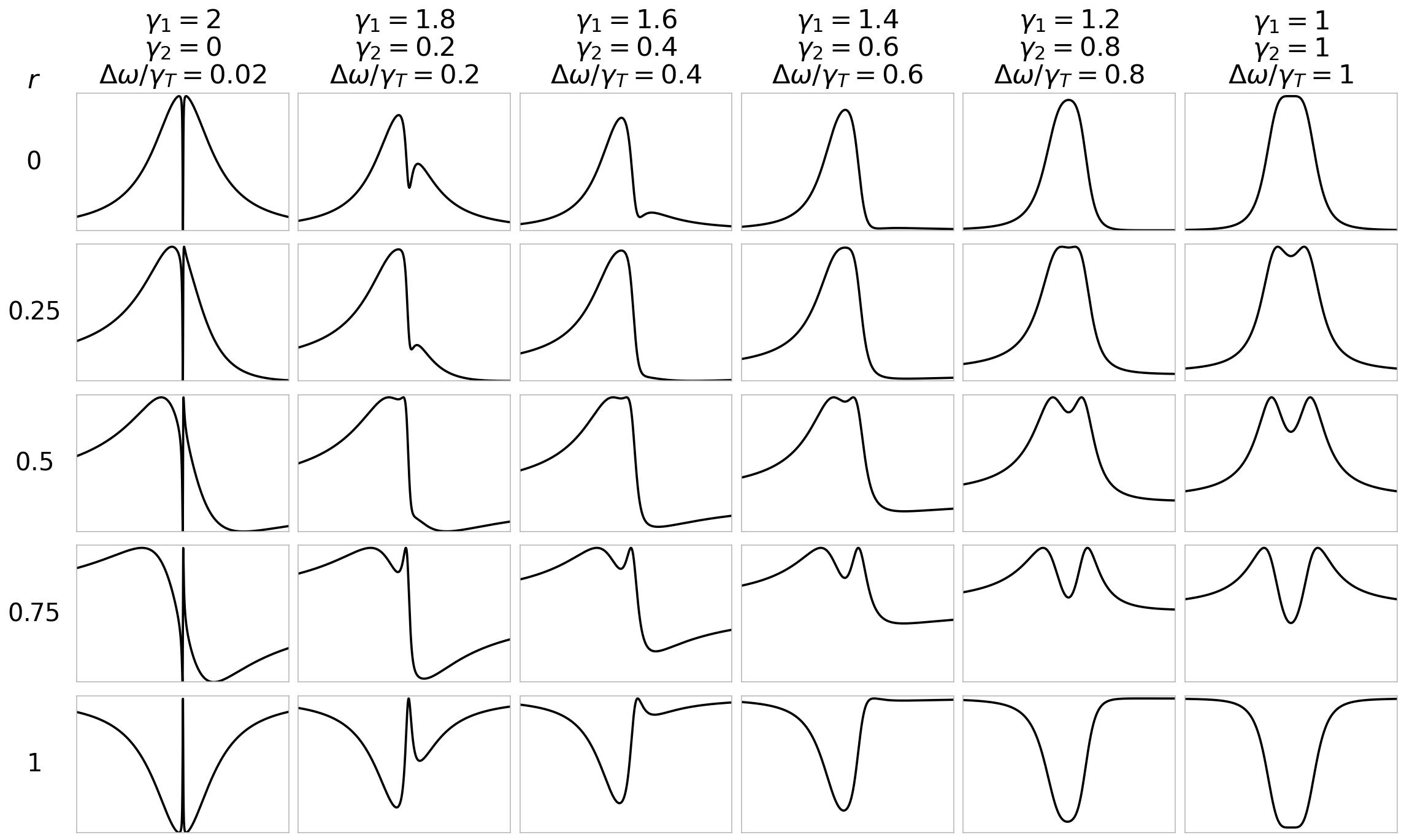}
    \caption{Intensity reflection spectra in a two-port, two-mode system as one varies the background amplitude reflection coefficient
$r$, decay rates $\gamma_1,\gamma_2$ and the frequency splitting  $\Delta\omega$. We choose $\gamma_T$ = $\gamma_1$ + $\gamma_2 = 2$. In each panel, the vertical axis is
the intensity with values ranging from 0 to 1. The horizontal axis is the normalized frequency
detuning of $(\omega-\omega_0)/\gamma_T$ with values ranging from -3 to 3. The other parameters are chosen to have a value of $\kappa = 0.2, \ \beta = 0, \ \omega_0 = 0$.}
    \label{fig:2modesweep}
\end{figure}

In Fig. \ref{fig:2modesweep} we illustrate
 a set of lineshapes from Eq. \eqref{eq:2modeS}. We choose $\kappa = 0.2, \ \beta = 0, \ \omega_0 = 0$  for all panels in Figure 3. Across each row, $\gamma_2$ increases from 0 to 1 while $\gamma_1$ decreases from 2 to 1 and $\gamma_T = \gamma_1 + \gamma_2 = 2$ remains constant. Further, the normalized detuning $\Delta\omega / \gamma_T$ increases from 0.02 to 1. These changes induce a transition from an electromagnetically induced transparency (EIT) spectrum to a flat-top spectrum. In the case of the EIT, the system has one dark mode, with $\gamma_2 = 0$; however, the nonzero $\kappa$ mixes the dark mode with the bright mode, allowing it to radiate through the radiation channel of the bright mode, thus producing the observed narrow transparency feature. Moving down the rows, we see that changing the background reflection coefficient $r$ changes the interference between the direct and resonant pathways, allowing for the emergence of a range of asymmetric lineshapes. Thus Fig. \ref{fig:2modesweep} illustrates a wide range of possible spectra that can be produced by our two-mode model. 
\section{\label{sec:level1}Numerical Validation of the Theory }
\begin{figure}
    \centering
    \includegraphics[width=0.5\linewidth]{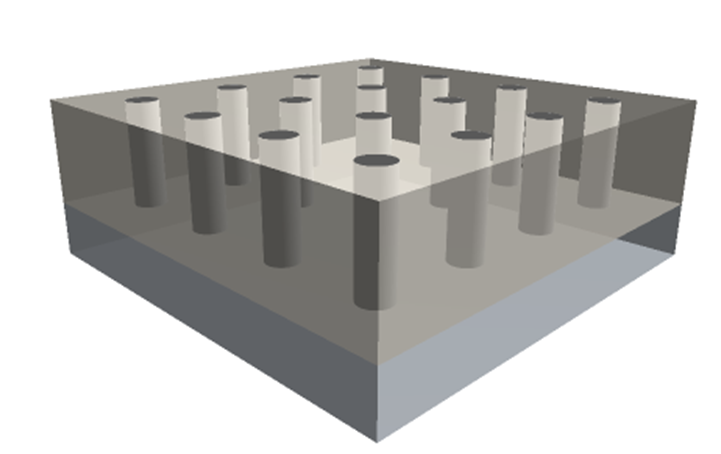}
    \caption{Schematic of a two-dimensional photonic crystal slab placed on top of a substrate. The photonic crystal has a square lattice of air holes with a lattice constant $a$ and a hole radius of $r_h$. The substrate is taken to be semi-infinite. The dielectric constants of the photonic crystal and substrate are $\epsilon = 12$ and $\epsilon = 4$, respectively. The region above the photonic crystal is assumed to be air. In the diagram, the thickness of the photonic crystal is taken to be $a$, with hole radius \\ $r_h$ = $0.15a$. }
    \label{fig:PHC}
\end{figure}

To validate our theoretical analysis, we compare the theoretical results derived in Section II to rigorous coupled wave analysis (RCWA)  simulations of a photonic crystal slab supporting resonant scattering features. For our demonstration, we choose our setup to be a photonic crystal slab placed on top of a uniform substrate (Fig. \ref{fig:PHC}). The substrate is assumed to be semi-infinite. The photonic crystal slab consists of a square lattice of air holes introduced into a high-index dielectric slab.    We simulate the transmission and reflection spectra in the normal direction for linearly-polarized incident light. Due to the symmetry of the structure, the polarization is preserved in the transmission and reflection processes. With the periodicity chosen to be smaller than the operating wavelength of light in the substrate, such a system naturally realizes a two-port system at normal incidence, with one port corresponding to the air region above and one port corresponding to the substrate region below the photonic crystal slab. In this system, the vertical mirror symmetry of the system is broken by the presence of the substrate, allowing the radiative coupling into the two ports to become unequal. We choose such asymmetric structures since our theory does not assume any structural symmetry.  

\begin{figure}
    \centering
    \includegraphics[width=0.6\linewidth]{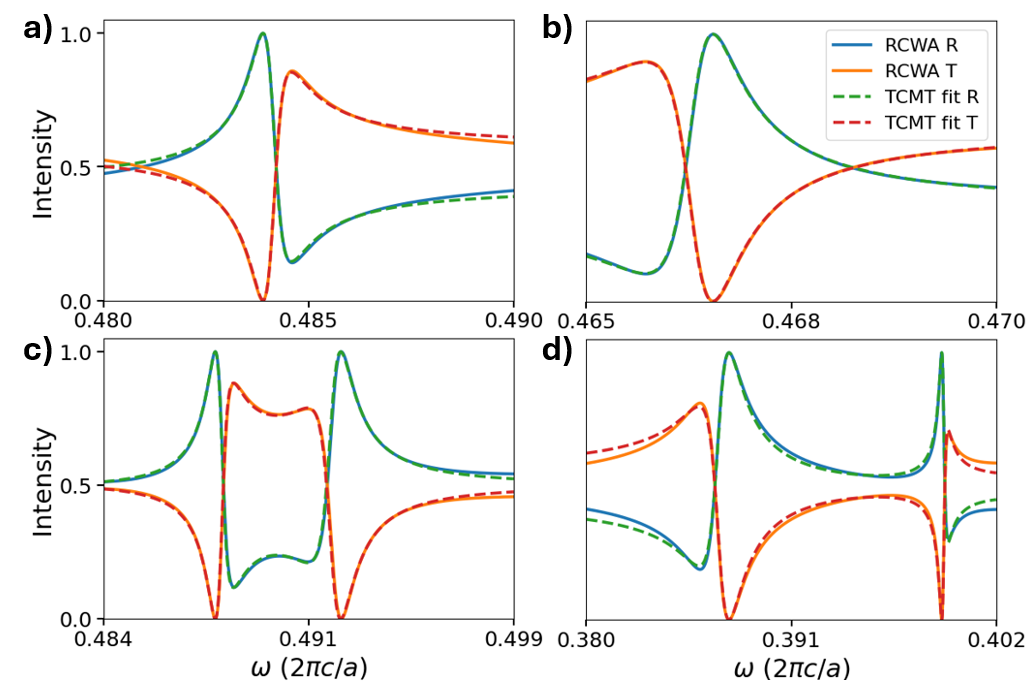}
    \caption{Fit of our TCMT model to RCWA spectra. Panels \textbf{a}) and \textbf{b}) show the single-mode case. Panels \textbf{c}) and \textbf{d}) show the two-mode case. The geometry and the fitting parameters are provided in Table \ref{tab:tcmt_fit_parameters}. }
    \label{fig:fitplots}
\end{figure}

In Fig. \ref{fig:fitplots} we fit our RCWA-generated spectra to our theoretical descriptions. The geometric parameters and frequency range of our system are chosen to produce some of the nontrivial lineshapes that our theory predicts. In the single-mode case, the RCWA spectra are fit using the two-port, single-resonance expression found in Eq. \eqref{eq:S_twoport}, where the fitted parameters are $r, \alpha, \gamma, \omega_0$. In the two-mode cases, the fitting parameters in addition include $\Delta\omega, \gamma_2, \kappa$, as well as $\beta$ instead of $\alpha$, as described in Eqs. \eqref{eq:2portomega} - \eqref{eq:p_2_final} with the scattering matrix from Eq. \eqref{eq:2modeS} used for fitting.

In all four of the cases presented in Fig. \ref{fig:fitplots}, with the geometry and the parameters of the coupled mode theory model provided in Table \ref{tab:tcmt_fit_parameters}, the TCMT model reproduces the spectral features with high fidelity. This agreement confirms that the scattering response in our photonic crystal system is well described as the interference between a slowly varying direct background pathway and the resonant pathways, with the parameters of the resonant pathway constrained by the background scattering matrix. Panels (a,b) correspond to the single-mode regime. In these cases, the spectra are dominated by a single isolated resonance, and our single-mode theory accurately produces the observed lineshape, and our fitted parameters are sufficient to capture the asymmetry of the resonant lineshapes. Panels (c,d) correspond to the two-mode regime, where two nearby resonances contribute to the spectra within the same spectral range. Here, we observe more complex spectral responses due to the superposition of the resonant contributions. Despite this, we are able to fit these results using our TCMT formalism to a similar level of precision as the single-mode case, indicating that our model can accurately capture the essential interference physics in the multimode case.

\begin{table*}
\caption{Geometric parameters of the structures and fitted TCMT parameters.}
\label{tab:tcmt_fit_parameters}

\centering
\begingroup
\small
\setlength{\tabcolsep}{2.5pt}
\renewcommand{\arraystretch}{1.08}
\newcommand{\tcmtHeader}[1]{%
  \begin{tabular}[t]{@{}c@{}}#1\end{tabular}%
}

\begin{tabular}{c|c|c|c|cc|cc|c|c}
\hline\hline
\tcmtHeader{Panel}
& \tcmtHeader{Case}
& \tcmtHeader{$r_h$\\$(a)$}
& \tcmtHeader{$r$}
& \tcmtHeader{$\omega_0$\\$(2\pi c/a)$}
& \tcmtHeader{$\Delta\omega$\\$(10^{-3}2\pi c/a)$}
& \tcmtHeader{$\gamma_1$\\$(10^{-3}2\pi c/a)$}
& \tcmtHeader{$\gamma_2$\\$(10^{-3}2\pi c/a)$}
& \tcmtHeader{$\alpha/\pi$ (1-mode)\\
              $\beta/\pi$ (2-mode)}
& \tcmtHeader{$\kappa$\\$(10^{-3}2\pi c/a)$} \\
\hline

\textbf{a)}
& Single-mode
& 0.150
& 0.658
& 0.484
& --
& 0.327
& --
& 0.901
& -- \\

\textbf{b)}
& Single-mode
& 0.150
& 0.585
& 0.466
& --
& 0.365
& --
& 0.593
& -- \\

\textbf{c)}
& Two-mode
& 0.200
& 0.703
& 0.490
& 3.508
& 0.292
& 0.523
& 0.183
& 1.017 \\

\textbf{d)}
& Two-mode
& 0.150
& 0.667
& 0.393
& 11.965
& 0.755
& 0.149
& 0.389
& 0.589 \\

\hline\hline
\end{tabular}

\endgroup
\end{table*}

The small deviations found in panels (a,c) between RCWA data and the TCMT model arise primarily from the fact that the background response in our RCWA simulation is not perfectly constant over our frequency range. In the analytic TCMT expressions we used to fit, the background scattering matrix is treated as a frequency-independent feature across the narrow spectral range in which our resonant features lie. This approximation is well justified when the background is slowly varying, but it is not exact in numerical simulation. Overall, the agreement between our RCWA simulations and the TCMT fit across all four examples provides strong evidence that the present formulation captures the relevant resonant scattering physics in photonic structures. 
\section{\label{sec:level1}Conclusion }
In this work, we presented a general construction for a reciprocal, lossless, temporal coupled-mode theory based on the Takagi factorization of the background scattering matrix. Our formulation explicitly constructs the allowed resonant coupling vectors through constraints imposed by the direct scattering channel, providing a compact parameterization of all single-mode resonance solutions for a fixed background.  Beyond the single-mode case, our formalism naturally extends to the multimode case by taking the resonant amplitude to be a vector and the coupling coefficients to be a port-by-mode dimensional matrix. This extension highlights the flexibility of our approach. We expect this construction to be useful for both analytical studies involving resonant scattering and for model fitting in nanophotonic and wave-based systems. More broadly, this construction provides a compact framework for characterizing the space of physically admissible resonant scattering responses under the constraints of reciprocity and energy conservation. 

\section*{Data Availability}
Data underlying the figures presented in this paper are not publicly
available at this time but may be obtained from the authors upon
reasonable request.

\section*{AI Disclosure}
GPT 6.1 Sol was used in the editing process of this manuscript. All changes and modifications suggested by the model were human-verified prior to implementation.

\begin{acknowledgments}
This work is supported by the Army Research Office (Grant No. W911NF-24-2-0170), and by a Simons Investigator in Physics grant from the Simons Foundation (Grant No. 827065). M.B. is supported by a Stanford Graduate Fellowship. S. F. acknowledges discussions with Cheng Guo on Takagi decomposition. 
\end{acknowledgments}

\bibliography{refrences}

\end{document}